\documentclass[sigconf,screen,nonacm]{acmart}
\usepackage{listings}
\usepackage{xcolor}
\usepackage{subcaption}
\usepackage{adjustbox}
\usepackage{tikz}
\usetikzlibrary{positioning, arrows.meta, shapes, calc}
\usepackage{multirow}
\usepackage{tablefootnote}
\usepackage{colortbl}
\usepackage{booktabs}
\usepackage[flushend]{}
\definecolor{codegray}{gray}{0.9}
\definecolor{aeroBlue}{HTML}{7CB9E8}

\lstdefinelanguage{customC}{
  language=Rust,
  morecomment=[l]{//},
  morecomment=[s]{/*}{*/},
  moredelim=**[is][\color{orange}\bfseries\Large]{@}{@},
}

\lstdefinestyle{mystyle}{
    language=C,
    backgroundcolor=\color{codegray},   
    commentstyle=\color{green},
    keywordstyle=\color{blue},
    numberstyle=\tiny\color{gray},
    stringstyle=\color{red},
    basicstyle=\ttfamily\scriptsize,
    breakatwhitespace=false,         
    breaklines=true,                 
    captionpos=b,                    
    keepspaces=true,  
    numbers=none,
    numbersep=5pt,                  
    showspaces=false,                
    showstringspaces=false,
    showtabs=false,                  
    tabsize=2,
    commentstyle=\color{green!60!black}\footnotesize\bfseries
}

\usepackage{textgreek} 
\usepackage{graphicx}       
\usepackage{balance}        
\usepackage{supertabular}   
\usepackage{array}          
\usepackage{float}          
\usepackage{enumitem}       
\usepackage{wrapfig}
\newlength{\figWidth}
\usepackage{placeins}
\usepackage{cuted}        
\usepackage{caption}
\usepackage[unicode]{hyperref}

\title{vApps: Verifiable Applications at Internet Scale}

\author{%
	Isaac Zhang\textsuperscript{*}, Kshitij Kulkarni\textsuperscript{†}, Tan Li\textsuperscript{*}, Daniel Wong\textsuperscript{*}, Thomas Kim\textsuperscript{*}\\
	John Guibas\textsuperscript{†}, Uma Roy\textsuperscript{†}, Bryan Pellegrino\textsuperscript{*}, Ryan Zarick\textsuperscript{*}%
}

\affiliation{%
	\institution{%
		\textsuperscript{*}LayerZero Labs \quad
		\textsuperscript{†}Succinct \city{} \country{}
	}%
}

\begin{document}

\begin{abstract}  
Blockchain technology promises a decentralized, trustless, and interoperable infrastructure. However, widespread adoption remains hindered by issues such as limited scalability, high transaction costs, and the complexity of maintaining coherent verification logic across different blockchain layers.
This paper introduces Verifiable Applications (vApps), a novel development framework designed to streamline the creation and deployment of verifiable blockchain computing applications. vApps offer a unified Rust-based Domain-Specific Language (DSL) within a comprehensive SDK, featuring modular abstractions for verification, proof generation, and inter-chain connectivity. This eases the developer’s burden in securing diverse software components, allowing them to focus on application logic. The DSL also ensures that applications can automatically take advantage of specialized precompiles and hardware acceleration to achieve consistently high performance with minimal developer effort, as demonstrated by benchmark results for zero-knowledge virtual machines (zkVMs). Experiments show that native Rust execution eliminates interpretation overhead, delivering up to an 197x cycle count improvement compared to EVM-based approaches.  Precompiled circuits can accelerate proving by more than 95\%, while GPU acceleration increases throughput by up to 30x and recursion compresses the proof size by up to 230x, enabling succinct and efficient verification. The framework also supports seamless integration with Web2 and Web3 systems, enabling developers to focus solely on their application logic. Through modular architecture, robust security guarantees, and composability, vApps pave the way toward a trust-minimized and verifiable Internet-scale application environment.
\end{abstract}

\hypersetup{pdfauthor={Layer Research}, pdftitle={Verifiable Applications}, pdfencoding=auto, unicode=true, pdfcreator={PDFLaTeX}, pdfproducer={LaTeX}, breaklinks=true, pdfescapeform=true}

\maketitle

\pagestyle{plain} 
\section{Introduction}
The grand vision for blockchain technology is a trustless, scalable, and interoperable digital infrastructure, enabling users and developers to interact with and deploy decentralized applications (dApps) across multiple blockchain layers. However, today’s prominent Layer 1 (L1) blockchains, such as Ethereum, continue to grapple with critical issues, including limited capacity, high transaction costs, and network congestion.
    
Current blockchain scaling strategies generally fall into two complementary categories: (1) monolithic: enhancing the L1 blockchain's inherent capabilities directly, and (2) modular: offloading computation to Layer 2 (L2) blockchains. In this model, the primary role of the L1 chain is settlement, maintenance of ledger data, and persistence of state commitments. Throughout this paper, the L1 chain will be designated as the settlement layer.  

The proliferation of rollups~\cite{scaling} is driven by the emergence of various verification techniques that ensure the settlement layer can independently verify the integrity of state transitions occurring in the L2 execution layers. This architectural decoupling allows blockchains to scale transaction throughput by separating execution from settlement and data availability. This separation of concerns mirrors the Internet's architecture, where the core network handles data routing, and applications execute data storage and transformation at the scalable edges. Typically, rollup implementations achieve transaction throughput that is 10x to 100x higher than that of monolithic blockchains like Ethereum (around 30 TPS)~\cite{rollupsurvey}. This improvement provides a clear path toward a Web2-like user experience in terms of settlement speed and transaction throughput.

However, today’s application developers must navigate the complex and fragmented landscape of rollup development, which introduces significant technical challenges. Maintaining separate codebases for on-chain verification and off-chain backend services imposes substantial cognitive overhead and increases the risk of security vulnerabilities. Additionally, the lack of a unified architecture leads to duplicated or inconsistent state across rollups, interconnected rollups, and Web2 components. This fragmentation expands the security surface and requires careful coordination and auditing across multiple layers, increasing development complexity.

In this paper, we introduce Verifiable Applications (vApps), a development framework that allows developers to focus solely on application logic without handling low-level concerns like transaction verification, data commitment, and connectivity that are common to all decentralized applications. vApps provide a unified development environment that abstracts proof generation, ensures secure inter-chain connectivity, and handles transaction submission/sequencing. Realized through the vApp SDK and its interoperable Rust DSL, the framework not only addresses codebase fragmentation but also tightly guards the transaction lifecycle to ensure code execution integrity and facilitate downstream verification. 

vApps are made possible by advances in verifiable computing techniques that enable off-chain layers to efficiently prove the integrity of transaction execution to the settlement layer. While originally restricted to optimistic verification techniques, recent innovations in general purpose zero-knowledge virtual machines (zkVMs) make verifiable computing a central pillar of the vApp paradigm. zkVMs can prove the execution of arbitrary code without incurring the overhead of specialized virtual machines like the EVM, as we show through a comprehensive study. 

vApps abstract the complexities of proof systems and Web3 domain expertise, empowering Web2 developers to create verifiable decentralized applications using only Rust and their existing domain models. This approach promises to accelerate Web2 adoption of decentralized technologies. Ultimately, vApps enable fully trustless deployment, reducing dependence on central authorities to coordinate execution across blockchains and Web2 systems. Their ability to verify data and computation across diverse environments will position them as a key layer for secure Internet applications and hybrid Web2-Web3 solutions. 

\section{Background}
Blockchain technology was initially envisioned to enable decentralized, trustless systems for immutable record-keeping, reducing the reliance on centralized authorities while also empowering individuals with sovereign control over their assets through permissionless access. This vision spurred the development of platforms such as Bitcoin and Ethereum. However, early blockchain designs followed a monolithic architecture, where data availability, consensus, execution, and settlement were all tightly integrated and executed by every node. Although this approach ensured strong security and fault tolerance, it introduced significant scalability challenges. Limited transaction throughput and rising costs under growing network demand have restricted broader adoption, especially for applications that aim to deliver a user experience comparable to Web2 services.
\begin{figure}
		\centering
         \setlength{\figWidth}{0.43\textwidth}
		\includegraphics[width=\figWidth]{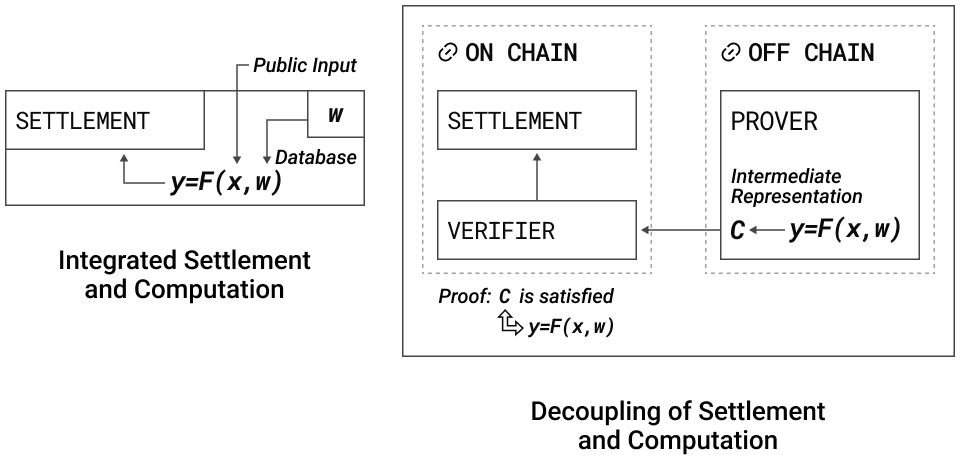}
        \Description{Description of the image goes here.}
		\caption{Decoupling settlement and logic execution by verifiable computing. Here, $x$ represents public input, $w$ denotes private data or witness, and $y$ is the resulting output of the computation $F(x,w)$.}
        \label{fig:modular}
	\end{figure}
Verifiable computing, as illustrated in Figure~\ref{fig:modular},  facilitates the transition toward modular blockchain architectures by decoupling computation from settlement. It enables trustless interaction between components and ensures the correctness of external computations without requiring full re-execution. These properties collectively address scalability limitations while preserving the security and decentralization of the base layer. This decoupling enables specialization and optimization across the stack and lays the foundation for building internet-scale applications.

\subsection{Rollups}

   Blockchain systems broadly consist of three core components: an execution layer for processing application logic, a data availability layer for maintaining persistent transaction records, and a settlement layer that finalizes and canonicalizes state through consensus. These protocols are inherently designed to resist Byzantine threats, including data tampering, historical revision, and transaction censorship.
	
\begin{figure}
		\centering 
        \setlength{\figWidth}{0.42\textwidth}
		\includegraphics[width=\figWidth]{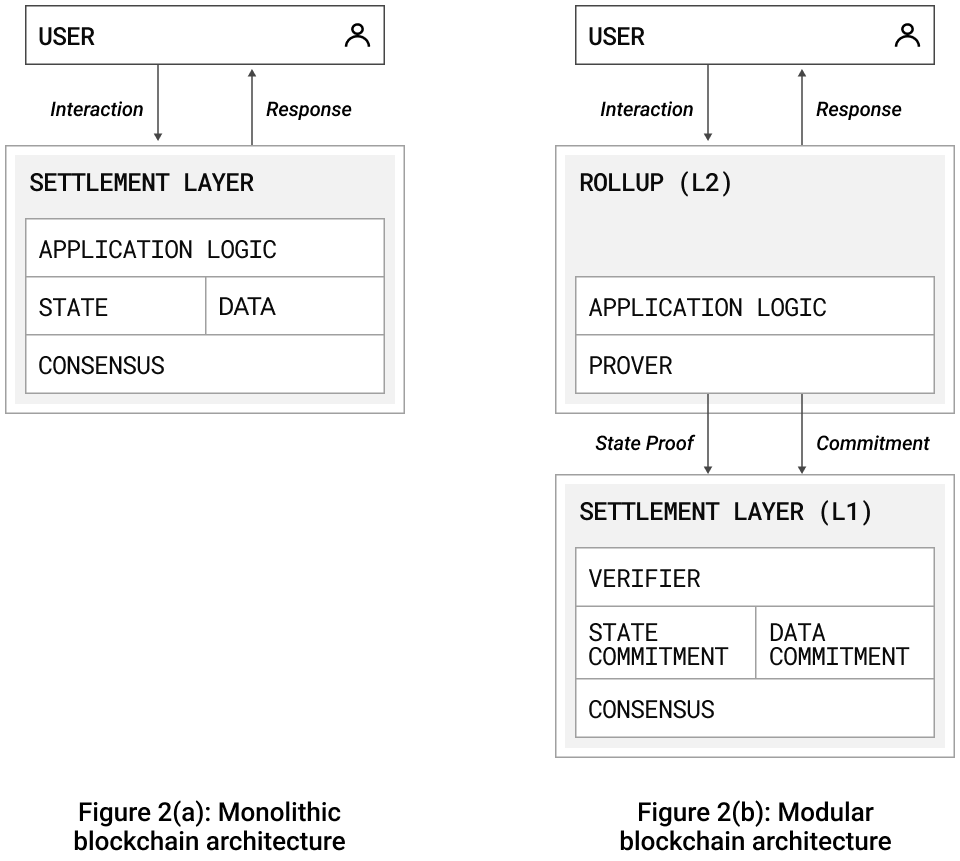}
            
			\Description{Comparison of monolithic and modular blockchain architectures.}
		  \caption{Comparison of monolithic and modular blockchain architectures.}
		\label{fig:archcomparison}
	\end{figure}

Figure~\ref{fig:archcomparison}(a) illustrates a monolithic blockchain architecture, where tightly integrated layers execute application logic and consensus within a single settlement layer. This consolidation simplifies design but leads to scalability limitations, as resource contention and computational overhead are concentrated in one layer.
	 As shown in Figure~\ref{fig:archcomparison}(b), modular blockchain architecture separates the execution layer from the settlement layer. Users interact with application logic on the L2 layer, where a Prover generates an integrity proof for each batch of transactions. This proof is submitted to the settlement layer to verify correctness and challenge invalid or censored transactions. The settlement layer provides security for proof verification, data commitment, and state finality. Offloading execution to the L2 reduces the L1’s computational burden, addressing the scalability limits of monolithic designs. The modular architecture can extend execution layers beyond simple transaction processing. For instance, emerging environments such as the vApp SDK can support general-purpose computation or data processing, going beyond typical rollup use cases. 

\subsection{Verifiable Computing}
Verifiable computing is a model in which a Prover executes a function $y=F(x,w)$ and produces a succinct proof attesting that the resulting output $y$ is correct, given a program $F$, public inputs $x$, and a witness $w$, as shown in Figure \ref{fig:modular}. A Verifier then checks this succinct proof, confirming it without having to re-run the function. Verifiable computing techniques rely on various trust assumptions and may use software-based or hardware-based attestation. 
The three types of verifiable computing techniques are:

\begin{itemize}[leftmargin=0pt,label={}]
    \item \textbf{(1) Optimistic:} This technique optimistically executes transactions off-chain and posts state updates without immediate proofs. Watchers can challenge potentially fraudulent transactions by submitting fraud proofs during a defined period. On the settlement layer, fraud proofs are verified, and invalid updates can be reverted. The trust assumption is that there is at least 1 honest and incentivized watcher who will monitor the rollup and submit fraud proofs when necessary. The complexity of the code used to verify fraud proofs introduces a significant risk, and its large and intricate nature increases the likelihood of bugs.
    
    \item \textbf{(2) Zero Knowledge (ZK) Proofs:} This technique uses cryptographic (zk-SNARK~\cite{snark} or zk-STARK~\cite{stark}) methods to generate succinct off-chain attestations of correct computations that produce state updates. These succinct proofs are then efficiently verified on-chain without dispute. The trust assumption is in the soundness of the underlying ZK proofs and the correctness of their implementation. Recent advancements in zero knowledge virtual machines (zkVMs) have made it possible to integrate ZK proof verification in rollup and decentralized application development at scale. Section~\ref{chap:zkvm} provides a deeper technical exposition and empirical study of this promising approach.
    
    \item \textbf{(3) Trusted Execution Environments (TEEs):} This hardware-based technique utilizes CPU or GPU enclaves to create a confidential computing environment that is isolated from the influence of the underlying operating system. This hardware attestation process can show an on-chain verifier that the code execution and state updates that happen in the TEE~\cite{cryptoTEE} are correct. The primary trust assumption is in the integrity and security of the hardware provided by the TEE manufacturer.   
    
\end{itemize}

 \begin{figure}
		\centering
        \setlength{\figWidth}{0.46\textwidth}
		\includegraphics[width=\figWidth]{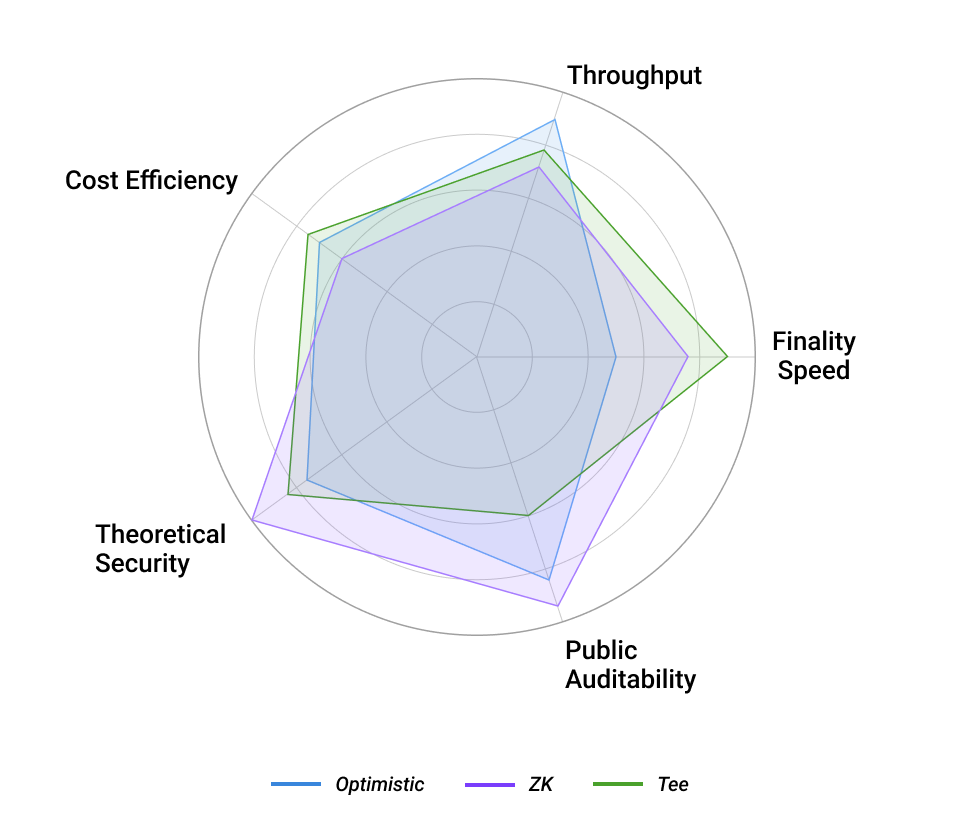}
        \Description{Description of the image goes here.}
		\caption{Comparison of trade-offs between different verifiable computing techniques.}
    \label{fig:radarChart}
	\end{figure}
Rollup solutions, built on the previously discussed verifiable computing techniques, support diverse application requirements according to each technique's performance and security trade-offs, as shown in Figure~\ref{fig:radarChart}.

Optimistic rollups~\cite{optrollup} have gained wide adoption and offer cost-effective deployments. They provide the highest throughput at low costs due to batch processing and deferred verification. However, the dispute challenge period (up to 7 days) reduces finality speed relative to ZK rollups and TEEs, which reduces suitability for rapid settlement. They also require accessible data for fraud proofs, and alternate data availability approaches that can weaken theoretical security.

When the highest level of trust-minimized verification is required, ZK rollups are the ideal choice because they use auditable cryptographic methods, such as zero-knowledge proofs, to verify the integrity of every transaction in batches. They are often used for cross-chain applications like transferring assets or verifying states across different blockchains. However, their higher computational overhead increases costs and reduces throughput. Due to the open sourcing of the underlying ZK proof systems, ZK rollups can benefit from audits and increasingly from formal verification methods.

Finally, TEEs are well-suited for lightweight rollup deployments that require low-cost off-chain operations, high throughput, and fast finality. Although they provide fast finality through efficient hardware pipelines, their proprietary nature means that any security vulnerabilities must be addressed through firmware updates from the manufacturer, resulting in lower public auditability compared to other solutions.

\section{Vapp Overview}
Despite rollup advancements, the absence of standardized architectural frameworks has resulted in significant codebase fragmentation. For example, zkEVM~\cite{codeFrag} implementations use heterogeneous components and languages, which complicates full verification and integration, undermining the consistency of contract code, the underlying infrastructure, and proof logic. This fragmentation is further compounded across different rollup architectures, amplifying incompatibility and integration complexity. Moreover, the requirement for deep blockchain module expertise hinders adoption by Web2 developers. The advent of verifiable computing provides a unique opportunity for upgrading the modular blockchain stack by abstracting the intricacies of Web3 and boosting developer adoption. 

\subsection{Verifiable Computing Made Simple}
\begin{figure}[ht]
		\centering
        \setlength{\figWidth}{0.42\textwidth}
		\includegraphics[width=\figWidth]{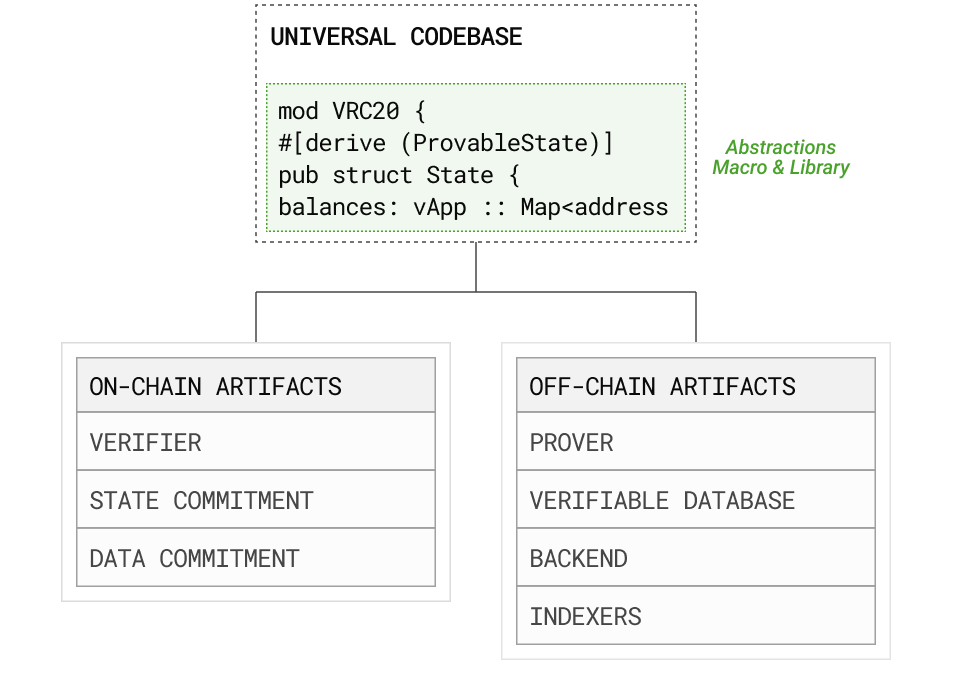}
        \Description{Description of the image goes here.}
		\caption{Hierarchical view of the generation of different application artifacts from vApp's universal codebase.}
    \label{fig:Vsimple}
	\end{figure}

From the developer’s perspective, vApps simplify verifiable computing applications by providing a hierarchical development model built around the vApp SDK. As shown in Figure~\ref{fig:Vsimple}, the SDK offers a universal codebase, allowing a \textit{single development environment} to automatically generate both on-chain and off-chain artifacts. 

This unified framework eliminates the need to maintain separate programming environments for verifiers, provers, and backend systems. As a result, developers can build verifiable computing applications more efficiently and reliably, avoiding the complexity typical in zkEVM development and reducing code fragmentation across rollup stacks.

    \subsection{Unified Development Environment}

\begin{figure}[ht]
		\centering
        \setlength{\figWidth}{0.44\textwidth}
		\includegraphics[width=\figWidth]{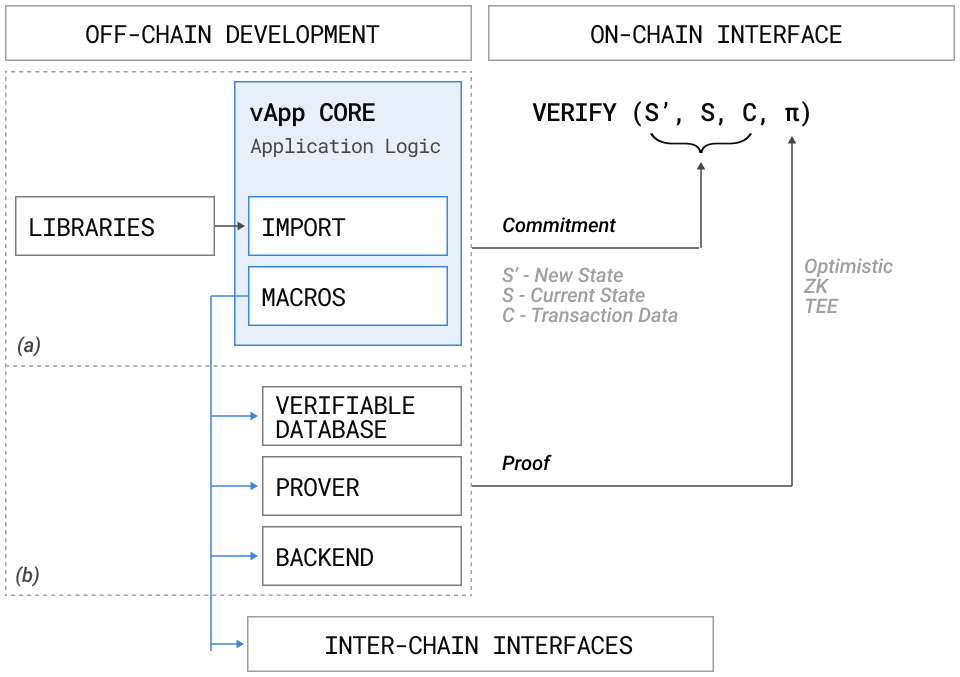}
        \Description{Description of the image goes here.}
		\caption{Illustration of the interaction between different layers supported by the Unified Development Environment (a) The vApp computing core. (b) Expanding the computing core by integrating with other layers to ensure data integrity, computation integrity, and interoperability.}
    \label{fig:Vcore}
	\end{figure}

    Figure~\ref{fig:Vcore} illustrates the interaction between all major components in the vApp framework. Developers only need to focus on the colored unit—defining their application logic using imports and macros within a Unified Development Environment. vApps provide a general framework for building trust-minimized services, independent of the underlying scaling solution. While they can be deployed within rollup environments, they are not limited to rollup-based architectures. Developers retain full control over the state machine and the economic features of their applications. For instance, a vApp can implement gasless transactions or specify a custom asset, such as a stablecoin, as its gas token. Additionally, a vApp can also implement custom system-level tasks like a cron job, which is not supported on the existing rollups.
    
    Once the application is written, the system automatically generates the artifacts as depicted in Figure~\ref{fig:Vsimple}.   
    
    The Unified Developer Environment provides standardized interfaces across diverse blockchain environments:
    \begin{itemize}[leftmargin=0pt,label={}]
    \item \textbf{Unified Off-chain Development:} Figure~\ref{fig:Vcore}(a) illustrates the vApp core, which encapsulates application logic and defines verifiable state transitions using imports and macros, forming the minimal unit for building vApps. In Figure~\ref{fig:Vcore}(b), the role of the verifiable database is to maintain application state and ensure data integrity during both read and write operations. To strengthen trust in computation,
    the vApp core can be extended by integrating with a prover, as shown in Figure~\ref{fig:Vcore}(b). This extension forms the basis for optimistic or ZK rollup-style attestation of computation integrity on trustless hardware. In the case of TEEs, the same application logic is executed inside a trusted enclave. This method provides hardware-based attestation on code execution integrity.
  
Developers can define the above options using macro abstractions and focus only on customizing their application logic; we describe the structure of these macros in Section \ref{sec:vapp-dev-framework}. The system provides macros for execution verification, standard libraries for authentication and voting, and configuration options for data availability and settlement layers. If the defaults are insufficient, developers can extend these functionalities through clearly defined, standardized interfaces.
\\
    \item \textbf{Unified On-chain Interface:} 
    The on-chain contract receives the commitments—comprising the current state ($\mathbf{S}$), updated state ($\mathbf{S}^{\prime}$), and transaction data ($\mathbf{C}$), along with a proof ($\boldsymbol{\pi}$). A unified on-chain interface verifies the state transition ($\mathbf{S}$, $\mathbf{C}$) → $\mathbf{S}^{\prime}$ using the same verification function Verify($\mathbf{S}^{\prime}$, $\mathbf{S}$, $\mathbf{C}$, $\boldsymbol{\pi}$), regardless of whether the proof is generated via optimistic, ZK or TEE techniques.
    \\ 
    \item\textbf{Unified Inter-chain Interfaces:} Standardized messaging protocols like LayerZero \cite{layerzero} can eliminate the need to build custom connectivity solutions while ensuring compatibility and security. A unified communication layer enables vApps to perform seamless cross-chain transactions. It ensures consistency (atomicity) while integrating functionalities from different chains to support the creation of interconnected decentralized services (composability). As a result, atomic, composable vApps can be designed to operate in both an asynchronous and synchronous manner.

    \end{itemize} 

\subsection{vApp Endgame} 
\begin{figure}[ht]
		\centering
        \setlength{\figWidth}{0.49\textwidth}
        \includegraphics[width=\figWidth]{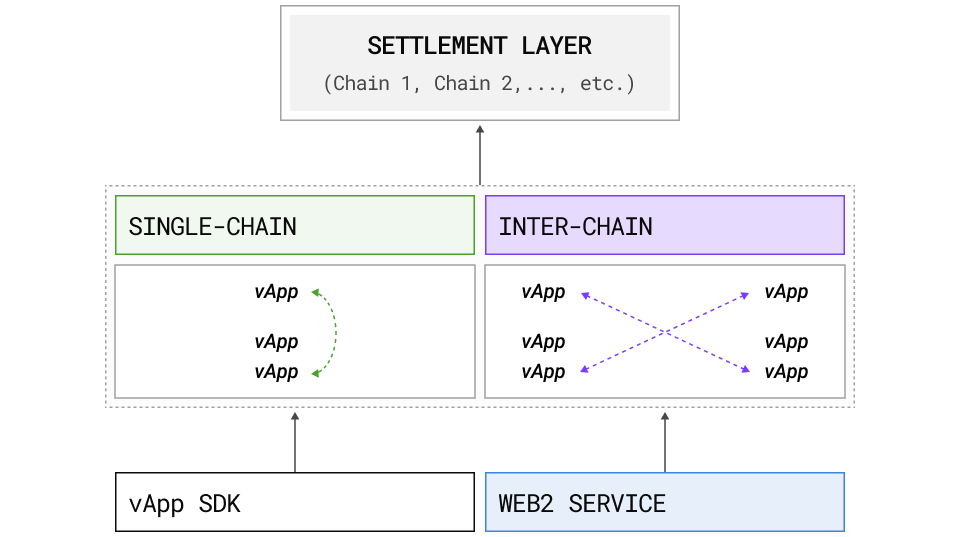}
	   \Description{An illustration of the vApp Worldview modular multi-chain architecture}
		\caption{The goal of the vApp framework is to provide a unified development environment for launching verifiable blockchain applications in single-chain, inter-chain, and Web2-Web3 hybrid scenarios.}
        \label{fig:vappVision}
    \end{figure}
    Figure~\ref{fig:vappVision} depicts a system designed to simplify the creation of vApps across multiple blockchain environments. At its core, the stack employs a modular architecture that abstracts the underlying complexity and streamlines developer workflows. The modularity also enables independent verifiable state transitions and consensus processes across distinct chains. Consequently, this structure achieves inter-chain interoperability while minimizing reliance on trusted intermediaries.

    The vApp SDK, implemented in Rust, provides high-level macros that abstract cryptographic proofs, such as zero-knowledge proofs, and chain-specific logic. This allows traditional Web2 services to integrate verifiable computing into existing applications without requiring developers to have detailed knowledge of cryptographic or blockchain primitives.
    
   Building upon the above-mentioned capabilities, this development framework supports key application scenarios, including:
    
    \begin{itemize}[leftmargin=0pt,label={}]
		\item \textbf{(1) Single-Chain Multi-vApp Settlement:} In scenarios limited to single-chain deployments, as depicted in Figure~\ref{fig:vappVision}, vApps achieve trust minimization by relying on the native blockchain consensus protocol and the verifiable database previously shown in Figure~\ref{fig:Vcore}(b). This model simplifies security assumptions and is optimal for applications prioritizing data integrity. Furthermore, since vApps support proof aggregation across multiple transactions -- as discussed in Section~\ref{chap:proofagg} -- they also enable atomic composability, allowing multiple vApps to be combined into a single cohesive application. Consider an atomic swap with intricate conditions that depend on multiple on-chain events or the state of different contracts within the same chain. Instead of verifying each condition individually, an aggregated ZK proof can validate them all at once. 
        \\ 
        
		\item \textbf{(2) Inter-Chain Multi-vApp Deployment:} Multi-chain vApps require reliable cross-chain data propagation facilitated by trusted modular data availability infrastructures. This model inherently introduces additional trust assumptions regarding inter-chain data availability and verification of events across chains. However, with trusted bridging protocols, verifiable inter-chain proofs, data integrity, and finality assurance, vApp applications can interact with a variety of blockchain features, liquidity pools, and user bases. For example, an inter-chain atomic swap ~\cite{atomSwap} could allow a user to exchange tokens on a DEX (Chain A) and deposit collateral in a lending vApp (Chain B), ensuring that either both operations succeed or neither does. 
        \\
       
		\item \textbf{(3) Merging Web2 and Web3:} With universal accessibility, as illustrated in Figure~\ref{fig:vappVision}, Web2 applications can tap into these ecosystems just as they leverage cloud services.
	vApps offer a universal on-chain address --- analogous to a Web2 IP address --- that ensures consistent and seamless access across different networks.
	For example, using vApp as a service, gaming applications that traditionally follow Web2 development practices can integrate verifiable computing and authentication
	without the need to learn an entirely new platform. This means developers can continue using familiar languages like Rust while adding secure, transparent, and verifiable blockchain features to manage device identity, data integrity, and access control. The  applications can choose to migrate the complete logic or only the parts that constitute credible commitments to their users. 
  \end{itemize}
    	
    \section{Architecture Overview}
\begin{figure}
		\centering
  
        \setlength{\figWidth}{0.47\textwidth}
		\includegraphics[width=\figWidth]{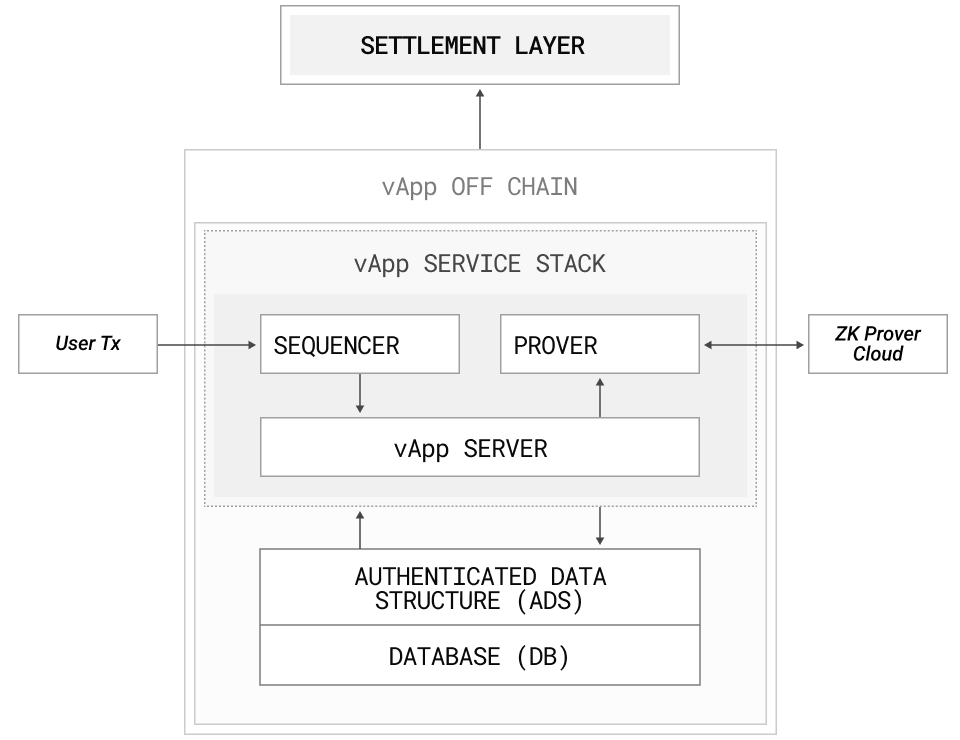}
		\caption{vApp Technical Stack Architecture: Interactions between off-chain execution, authenticated data storage, and L1 blockchain for settlement.}
		\Description{Architectural overview of interaction between the off-chain execution environment and the settlement layer.}
        \label{fig:vAppTechStack}
	\end{figure}

    Figure~\ref{fig:vAppTechStack} illustrates the architectural overview underlying the vApp framework. The architecture integrates service-oriented components (Sequencer, vApp Server, Prover) that interact with the Database and utilize Authenticated Data Structures to achieve reliable and secure single-chain and inter-chain applications and verifiable web2 applications. 
 
    \subsection{Service-Oriented Components}
    \label{chap:techStack}
    The modular components within the architecture, each performing specific functions, are described below.  
    
    \subsubsection{Sequencer.} Serves as the interface through which users can initiate actions and submit requests to the decentralized application. This submission interface authenticates transactions and orders them for processing in the vApp Server. It then publishes the ordered transaction data to a data availability layer so that external verifiers ("watchers") can independently confirm the correctness of sequencer behavior and the application logic. It can also provide cryptographically binding preconfirmation of transactions for users with extremely low latency.

    \subsubsection{vApp Server.}Serves as the orchestrator between the Sequencer, the data layer, the Prover, and the full nodes of the settlement layer. It supports the following operations:  
       
            \paragraph{Logic Execution.}Ensures that the appropriate verification method is initiated based on the configuration, and manages the full lifecycle of proof generation and verification, including dispute resolution in optimistic rollups.

            \paragraph{Connection to Settlement Layer.} 
            Establishes RPC (Remote Procedure Call) connections to full nodes of the relevant blockchain(s) to read the current on-chain state, retrieve transaction data required for proof generation, and monitor events. Light client protocols with Merkle proofs enforce data integrity, as any tampering by a full node results in rejection. Once computations are verified, the resulting compressed transaction data or state roots are submitted to the designated settlement layer for data availability and final settlement.

            \paragraph{Structured Data Indexing.}Automatically indexes transaction and state data based on macro-defined data structures. The resulting indexed data enables efficient off-chain querying via GraphQL~\cite{graphQLBook} by external applications such as data visualization dashboards and audit tools. 
            \paragraph{Cross-Chain Logic}Defines the dependencies and rules that govern transactions and state changes across multiple blockchains.
       
        \subsubsection{Prover}Generates cryptographic proofs, such as ZK proofs, and submits them to verifier contracts on the settlement layer. If these proofs are verified on-chain, they establish the validity of certain state transitions or data integrity without having to run the computation again. vApps provide user-configurable (see Figure~\ref{fig:Vcore}) options for software- and hardware-based attestation to generate computational integrity proofs. The option of hardware attestation can be implemented through a TEE-enabled Prover node, supported by specialized TEE hardware such as Intel SGX or ARM TrustZone.

	\subsubsection{Verifiable Database}
    As shown in Figure~\ref{fig:vAppTechStack}, the verifiable database, composed of the Database (DB) and Authenticated Data Structures (ADS), forms the bedrock of the system's operations. Conceptually, a verifiable database consists of these two components; however, in practice, they can be integrated into a unified layer, such as QMDB\cite{qmdb}.
  
     \paragraph{Database (DB)}
    Serves as the underlying datastore for managing the vApp's system state. It often utilizes high-performance key-value stores such as RocksDB or LevelDB that are specifically optimized for low-latency random access to handle frequent state transitions. This database also has witness data, which includes transaction payloads, execution trace logs, and records of state changes. The Prover uses this data to construct a proof attesting to the correctness and integrity of the transaction execution.

    \paragraph{Authenticated Data Structure (ADS)}Acts as an abstraction layer above the underlying database, enhancing it with the ability to generate verifiable evidence about its contents, known as data commitments. This allows external parties to trustlessly confirm the availability of specific information through data inclusion or exclusion proofs. Moreover, the ADS can send committed vApp state data to the settlement chain or other layers, thereby enabling trust-minimized verification of the vApp's state both on-chain and by other interested parties. This mechanism further facilitates lightweight validation of the vApp's state integrity during State Transition Function (STF) executions, as discussed in the next section.

	\subsection{Transaction Lifecycle}
    \label{chap:cycle}
	Ensuring the integrity of off-chain transaction execution is a cornerstone of vApp's architectural design.
    \begin{figure}[ht]
		\centering
        \setlength{\figWidth}{0.41\textwidth}
		\includegraphics[width=\figWidth]{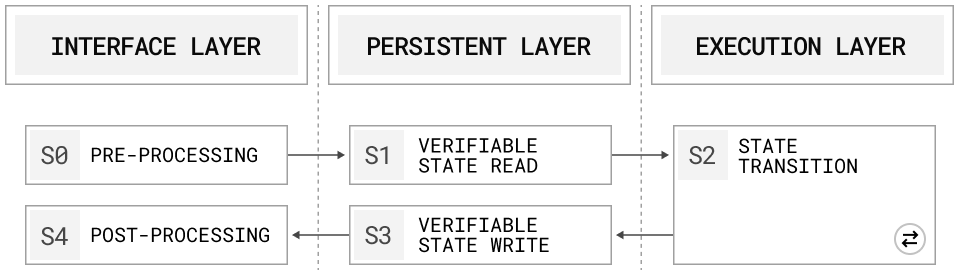}
		\caption{Step-by-step view of the transaction lifecycle.}
		\Description{Step-by-step view of the state transition function lifecycle, showing the process in detail.}
        \label{fig:vAppSTF}
	\end{figure}
   
	Figure~\ref{fig:vAppSTF} provides a more granular perspective on transaction execution:
 
    \begin{itemize}[leftmargin=0pt,label={}]
		\item \textbf{S0 (Pre-processing):} Performs checks on  $\mathbf{C}$, the transaction data (see Figure~\ref{fig:Vcore}) submitted by users, against required signatures, nonces, and format compliance, to filter out invalid transactions early.
        
		\item \textbf{S1 (Verifiable State Read):} State data is read from the ADS/DB using inclusion and exclusion proofs relative to $\mathbf{S}$, the current state of the vApp committed during the previous transaction cycle. This guarantees the integrity of the retrieved data.  
		\item \textbf{S2 (State Transition):} The state transition ($\mathbf{S}$, $\mathbf{C}$) → $\mathbf{S}^{\prime}$,  with memory tracking for a potential rollback, is executed by the State Transition Function. The new state is $\mathbf{S}^{\prime}$. 
		\item \textbf{S3 (Verifiable State Write):} $\mathbf{S}^{\prime}$ is committed to the ADS/DB, as the reference to the next transaction cycle.  
        \item \textbf{S4 (Post Processing):} Generation and storage of transaction receipts, event logs, and metadata for transparency and auditability in decentralized applications.
	\end{itemize}

	This phased classification defines clear verification points and cryptographic assurances (e.g., Merkle proofs) throughout transaction execution. By verifying input validity, state integrity, and outcome transparency, it safeguards trust in the entire process.

    \subsection{Execution Logs}
    \label{chap: Trace}
    The basis of verifiable computing is to prove that the execution of a computation—represented by its record of intermediate states and computational steps—adheres to a set of predefined rules or constraints. This ensures that a new state ($\mathbf{S}^{\prime}$)  is validly derived from the previous state ($\mathbf{S}$) through the application of a State Transition Function, as exemplified by the S2 phase discussed in Section~\ref{chap:cycle}. 

  \begin{table}[ht]
  \centering
  
  \Description{This table displays Execution Trace construction for one State Transition Function cycle. }
    \footnotesize
     \begin{tabular}{|>{\centering\arraybackslash}p{0.019\textwidth}|p{0.16\textwidth}|p{0.23\textwidth}|}
      \hline
    \textbf{Step}\hspace{0.8em} & \textbf{Action} & \textbf{Explanation} \\ \hline
      1 & Initial Log \newline (Log = [Transaction data, S]) & Given a starting state that is represented by S (e.g., a Merkle root). \\ \hline
      2 & Append Modifications \newline (Log = Log + [Sx, P(Sx)]) & Every intermediate state Sx and its commitment, P(Sx), is appended to the Log. \\ \hline
      3 & Compute New State Root \newline (S' = S union \{S1, S2, …\};\newline Log = Log + [S']) & Calculate the final state S' (i.e., the new state root) and append it to the Log by state commitment scheme  \\ \hline
      4 & Proof Submission & Send the log to the prover to generate the proof.  \\ \hline
    \end{tabular}
    \caption{Execution log construction for one state transition lifecycle.}
\label{tab:eTrace}
\end{table}
   Table~\ref{tab:eTrace} provides the formal specification for constructing an execution log compatible with structured program execution. It shows that a program starts from an initial state, and during its execution, a log of intermediate state changes ($S_{x}$) at each step is recorded. Associated with each of these steps, there might be a proof artifact ($P(S_{x})$), such as a Merkle proof demonstrating data inclusion within a Merkle tree. These are distinct from ZK proofs, which verify that all intermediate state changes leading to the final state are correct. Following these sequential modifications, the system arrives at a final state, which is often represented as a new state root.
    
     The execution log or witness is essential for the next step, proof generation. In the case of ZK, the prover then verifies if the collected witness adheres to all the constraints, confirming a valid execution. If the witness satisfies the constraints, the prover uses this information and the constraint system to produce a succinct zero-knowledge proof. More details can be found in Section~\ref{chap:zkvm}. 
  \subsection{Proof Aggregation: Composable vApps}
     \label{chap:proofagg}
     The modular architecture of vApps facilitates integration with modern cryptographic techniques like recursive~\cite{RecurZKP} SNARKs, STARKs, and proof-carrying data (PCD)~\cite{PCD}. These techniques enable the processing of multiple state updates by consolidating their execution logs or individual proofs into a single, concise aggregated cryptographic proof. A single verification of the aggregated proof on the blockchain ensures the integrity of multiple computations and enables composability of vApps within and across different blockchains. 
     
    Execution Log aggregation is most effective in homogeneous environments on a single chain, especially when applications operate under the same zKVM (see Section~\ref{chap:zkvm}). The combined constraint representation can capture the entire state transition triggered by the interaction of individual applications, so it allows a single proof for a whole batch of transactions. This approach can be simpler for smaller-scale, consistent, and well-coordinated operations, further highlighting the necessity for widespread adoption of vApps.
    
Conversely, recursive proof aggregation offers greater modularity and flexibility in heterogeneous, asynchronous environments with potentially different proving systems. It enables the creation of a combined ZK proof from applications without requiring a unified constraint system for all of them. It is generally more efficient for inter-chain composability.
	\section{vApp Development Framework}\label{sec:vapp-dev-framework}
	\begin{figure}[H]
		\centering
        \setlength{\figWidth}{0.45\textwidth}
		\includegraphics[width=\figWidth]{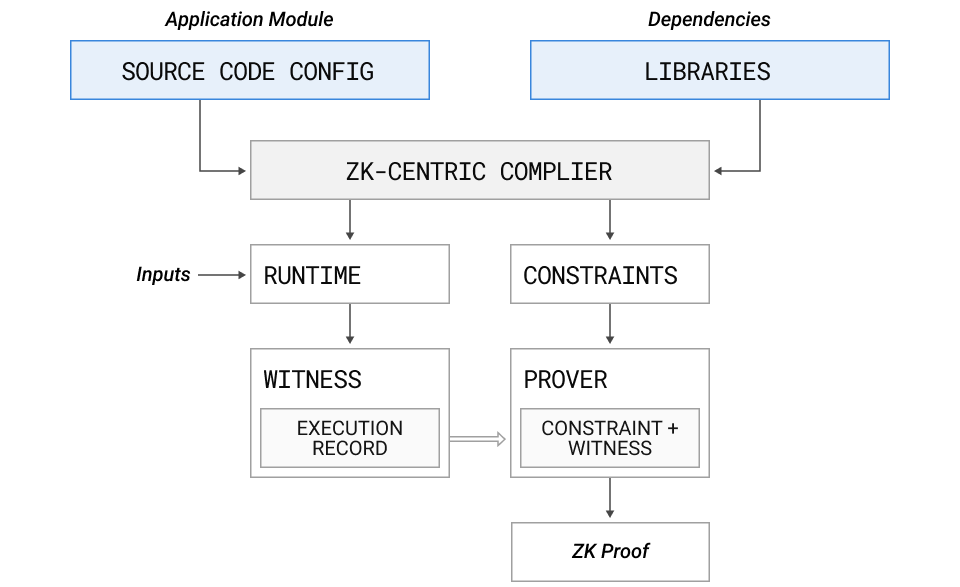}
		\Description{Diagram illustrating the vApp development and deployment framework.}
		\caption{Execution flow of a vApp project using ZK proofs.}
        \label{fig:vAppDev}
	\end{figure}

    \subsection{Modular Macro and Library Suite}
    Figure~\ref{fig:vAppDev} clarifies the boundary between application development, compilation, and proof generation, using the ZK case as an illustrative example. Developers initiate a vApp project by defining application logic and importing SDK libraries, as exemplified by the colored units. The SDK also provides macros to generate integrity proofs based on the application logic and can handle diverse computational processes beyond the limitations of the Ethereum Virtual Machine (EVM). These macros ensure that developers avoid subtle bugs or inefficiencies when working with cryptographic primitives.
    
    The program source code undergoes compilation via a ZK-centric compiler, yielding two primary artifacts: a runtime binary for the execution of the computation with concrete inputs, and a constraint system that is directly derived from the computation's logic. The execution engine, such as the zkVM introduced in Section~\ref{chap:zkvm}, produces an execution log or witness, which is then used by the Prover, along with the constraint system, to generate a succinct ZK proof.

    In addition, cryptographic primitives like BLS signature verification~\cite{blsSig} are also supported by macros. A developer can readily implement a mechanism that aggregates BLS signatures from multiple users to collectively authorize a state change, followed by ZK attestation of both the aggregated signature verification and the state transition.

    Similarly, on-chain settlement artifacts and inter-chain connectivity components are generated according to macro definitions in the code, ensuring that both verification and cross-chain communication are aligned with the application's declarative logic.
	
	\subsection{SDK Examples}
     A minimal version of a vApp in the Rust DSL is presented here as an illustration for following the steps S0-S4 of a program’s transaction lifecycle as discussed in Section~\ref{chap:cycle}. This practice facilitates the proof generation as introduced in Section~\ref{chap:zkvm} to benefit from specialized precompiles and hardware acceleration. The correspondence with the transaction lifecycle is highlighted through green-colored comments in the code script, as illustrated in Listing~\ref{lst:rustCode} below.

\begin{lstlisting}[numbers=left, caption={A minimal vApp example for a token transfer.},label={lst:rustCode}]
mod VRC20 {
    use super::*;

    pub type Address = [u8; 32];
    pub type Balance = u128;

    // Defining the domain model
    #[derive(ProvableState)]
    pub struct State {
        balances: vApp::Map<Address, Balance>,
    }

    #[derive(ProvableTx)]
    pub struct Transfer {
        from: Address,
        to: Address,
        amt: Balance,
    }
    
    #[derive(Event)]
    pub struct TransferEvent {
        from: Address,
        to: Address,
        amt: Balance,
    }

    #[vApp::Handler]
    pub fn transfer(tx: vApp::Context<Transfer> /* S0 Pre-processing */) -> Result<()> {
        /* S1 Verifiable state read */
        let from_balance = vApp::State().balances.get(tx.from).unwrap_or(0);
        let to_balance = vApp::State().balances.get(tx.to).unwrap_or(0);

        // Perform arithmetic and validation checks
        require!(tx.amt <= from_balance, ErrInsufficientFunds);
        require!(to_balance.checked_add(tx.amt).is_some(), ErrOverflow);

        /* S2 State transition & S3 Verifiable state write */
        vApp::State().balances.set(tx.from, from_balance - tx.amt);
        vApp::State().balances.set(tx.to, to_balance + tx.amt);

        /* S4 Post-processing */
        vApp::emit_event(TransferEvent {
            from: tx.from,
            to: tx.to,
            amt: tx.amt,
        });

        Ok(())
    }
}
\end{lstlisting}

\subsubsection{\textbf{Domain Definition}}
The on-chain ledger of user balances is represented by the \texttt{State} struct in 
Listing~\ref{lst:rustCode}, lines~7--11. Annotated with \texttt{\#[derive(ProvableState)]},
this ensures each balance update is captured in a commitment-friendly data structure.
By storing \texttt{(Address $\rightarrow$ Balance)} pairs in \texttt{vApp::Map}, the contract can generate cryptographic proofs (Merkle proof) of each individual balance change, to ensure that every state update, such as incrementing or decrementing an address’s balance, is provably linked to the corresponding transaction. 

Similarly, \texttt{\#[derive(ProvableTx)]} 
(lines~13--18) instructs the vApp execution log generator to produce a verifiable record of each transaction for a downstream Prover.
Finally, \texttt{\#[derive(Event)]} (lines~20--25) captures state transitions and other operations in a log file (\texttt{TransferEvent}), ensuring external audits can be done without reading low-level storage directly.

\subsubsection{\textbf{Transfer Logic}}
The core function that implements the token transfer logic is defined in  line~27 under \texttt{\#[vApp::Handler]}, designating it as the main 
transaction entry point. Below is a concise mapping of each step in the transaction lifecycle S0–S4, as discussed in Section~\ref{chap:cycle}, to the code in the token transfer logic:

\paragraph{S0 (Pre-processing).} Occurs when the transaction enters the transfer function, in line 28. 
\paragraph{S1 (Verifiable state read).}Within lines~30--31, the current balances for 
\texttt{tx.from} and \texttt{tx.to} are read from the on-chain state. 

\paragraph{S2 (State transition).} In lines~34--35, two checks ensure that (i) the sender has sufficient
funds, and (ii) the addition of \texttt{tx.amt} to the recipient’s balance will not 
overflow. Upon passing these checks, lines~38--39 apply 
the valid state transition by debiting the sender and crediting the recipient.
\paragraph{S3 (Verifiable state write).}As in lines 38–39, the new account balances are committed to an off-chain verifiable database. 
\paragraph{S4 (Post-processing).}In lines 42–46, the transaction details are logged by emitting a \texttt{TransferEvent}. This event aids debugging and off-chain audits, and it also forms part of the downstream proof generation process. On success, the function updates the on-chain state in a manner compatible with proof generation and returns \texttt{Ok(())} in line~48, indicating the transfer has completed.

\subsubsection{\textbf{Handling Execution Logs}}
	
	The execution record mechanism, as discussed in Section~\ref{chap: Trace}, ensures that every modification to the state is recorded as a witness for subsequent proof generation.
	To streamline this process, a \texttt{State()} object acts as an intermediary, automating the generation of inclusion proofs and updated state roots.
	Every state transition (S1, S2, ..., Sn) passes through the \texttt{State()} object, which enforces correctness and ensures that all modifications are properly logged.
	 This abstraction simplifies proof generation by automatically producing proof artifacts that link each new state back to its predecessor. This step simplifies the packaging of the execution record into the constraint system that the prover will attest to generate the proof of correct computation.

	After these steps, the proof and the new state root are sent to the settlement layer.
	If the proof is validated, the updated balances become part of the canonical on‐chain state.
	
	
	By encapsulating these tasks within the vApp SDK, developers are free to focus on core application logic without manually handling complex proof generation.

\section{zkVM Deep Dive}
\label{chap:zkvm}
\subsection{Overview}
Zero knowledge virtual machines (zkVMs) represent a \textit{general-purpose} and programmable approach to verifiable computation, implementing a virtual machine that can generate ZK proofs of arbitrary program execution. These systems abstract the underlying zero knowledge proof generation, allowing developers to write programs in high-level languages like Rust while outputting proofs that can be efficiently verified. They can be used to build scaling solutions for blockchains such as ZK rollups and enable general purpose verifiable computation for applications beyond blockchains. 

zkVMs operate at the assembly level, often by compiling source code down to open instruction sets like RISC-V. This provides significant generality in the kinds of programs that can be efficiently proven compared to EVM-centric approaches like zkEVMs, which operate at the bytecode level and aim to build scaling solutions that are equivalent to Ethereum \cite{buterin2022zkevm}. Because zkEVMs attempt to be compatible with the EVM or the full Ethereum protocol, which was not designed for ZK proving, they suffer from inefficiency, state access overhead, and significant developer penalties for writing \textit{application-specific} constraints of the proof system. zkVMs, in contrast, enable developers to write normal programs without incurring the overhead of EVM-specific opcodes and precompiles.

Modern zkVMs such as SP1, RiscZero, OpenVM, and Jolt allow developers to write programs in Rust, an unlock over less safe and expressive languages like Solidity \cite{succinct_sp1, risc0, openvm, a16z-jolt}. They can act as a central component of the vApp SDK by enabling a programmable, off-chain execution environment that is verifiable. They can support multiple execution environments across various blockchain ecosystems. This is a particularly beneficial feature for cross-chain applications such as DeFi platforms that operate on multiple blockchains. Today, zkVMs are being deployed in production applications, including in optimistic rollups that are converting to ZK rollups \cite{mantle2024zkrollup}, in bridging \cite{sp1-blobstream, sp1-vector}, and in novel multi-proof constructions \cite{taiko2024basedrollup}. 

\subsection{Architecture}
\begin{figure}[ht]
		\centering
        \setlength{\figWidth}{0.45\textwidth}
		\includegraphics[width=\figWidth]{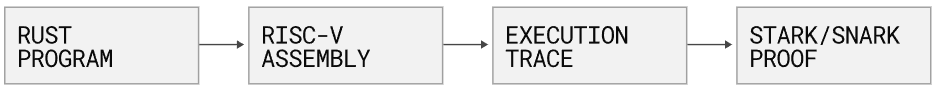}
		\Description{A schematic of the zkVM's workflow.}
		\caption{A schematic of the zkVM's workflow.}
        \label{fig:sp1-workflow}
	\end{figure}

We now describe the architecture of RISC-V zkVMs and highlight their highly modular and parallelizable nature, which makes them naturally suited to optimization at several levels of the stack. We describe how a program written in a high-level programming language is compiled down to RISC-V, executed, and proven by the zkVM. This process is depicted in Figure \ref{fig:sp1-workflow}.


\subsubsection{Rust program.} A developer writes the program to be proven in Rust. The developer can leverage Rust's rich ecosystem of crates while writing code that can be executed within the zkVM. This can include complex business logic, cryptographic operations, or operations on personally identifiable information that must remain private. 

\subsubsection{RISC-V Assembly} The developer compiles the Rust code into the RISC-V instruction set, which is a standard instruction set architecture (ISA). This compilation step generates a RISC-V ELF (executable and linkable format), which is a standard binary format for executable files. The number of RISC-V cycles it takes to execute the program is called the \textit{cycle count} and is often used as a measure of work for the proving system. The RISC-V ELF is then executed by the zkVM alongside some input to generate an execution trace that can be proven by the zkVM. 

\subsubsection{Execution Trace} The zkVM's runtime then uses the RISC-V ELF to process the program's data into an execution trace, which is a complete record of all the individual statements that need to be proven to ensure valid RISC-V execution. The execution trace is then converted into a set of prover traces that satisfy certain arithmetic constraints. These prover traces are utilized to generate a succinct proof, through proving protocols such as STARKs. Modern zkVMs can also be run in \textit{execution mode}, which only runs through the execution step on the RISC-V emulator, without proof generation. This is beneficial for program development and iteration because execution mode reveals critical details about the program's correctness and the amount of work necessary to generate a ZK proof, making zkVMs programs suitable for optimization.

\subsubsection{STARK/SNARK Proof} The proving system generates a STARK proof that attests to the correctness of the execution. The proof cryptographically verifies that the constraints, which implement a RISC-V virtual machine, were properly satisfied. The STARK proving process typically involves multiple phases, including polynomial commitments for committing to the execution trace, memory arguments, and a challenge step. The resulting proof can be publicly verified by anyone, confirming that the claimed computation was executed correctly according to the specified constraints. This proof can be further compressed into a SNARK, via the PlonK or Groth16 formats, which requires a trusted setup ahead of time \cite{gabizon2019plonk, groth2016size}. Whereas  na\"ive STARK verification on Ethereum incurs large gas costs, these formats are succinct and amenable to onchain verification. 

\subsection{Performance}
Due to their highly modular nature, zkVMs are suitable for optimizations at the proof system level and for hardware acceleration. In this section, we describe techniques for such optimization. First, we map the distribution of cycle counts to steps S0-S4 of the transaction lifecycle (Section~\ref{chap:cycle}) for a program that proves Ethereum execution with Reth. We then show how these cycle counts can be optimized via specialized precompiles for various cryptographic operations, how the EVM can be repriced to account for ZK proving costs, how zkVMs can avoid interpreter overhead with native code execution, and how hardware acceleration with GPUs can improve performance. We conduct these benchmarks on SP1 \cite{succinct_sp1}. 

    \subsubsection{Key Metrics.} The performance of zkVMs can be measured via latency, throughput, or proving cost.  We define these metrics and use them to profile the various optimizations we study in this section. 
    \paragraph{Latency.} The end-to-end time required to generate a ZK proof of a program is called the \textit{latency}. Typically, latency is measured in seconds or minutes, depending on the complexity of the program being proven and the efficiency of the proving system. Latency can be affected by factors such as the program's size, the underlying hardware, the zkVM implementation, and the chosen proving system. Lower latency means that the downstream application can verify the underlying transactions faster; this is especially critical for blockchain applications such as real-time proving in Ethereum, which aims to generate proofs of transaction execution within a single block time to provide fast finality guarantees to end users.
    
\paragraph{Throughput.} A measure of the work a zkVM does in a unit of time is the \textit{throughput}. Measured in KHz (thousands of cycles per second), throughput represents how many RISC-V cycles can be proven per second. Higher throughput indicates better performance and depends on factors such as hardware configuration, parallelization, and proving system optimizations. Throughput and latency are generally uncoupled and should be considered as independent metrics. Importantly, high throughput can be achieved even with high latency, especially in pipelined or parallelized settings. For instance, if a single machine takes 10 seconds to prove a 1-second block, the latency is 10 seconds and the throughput is 0.1 blocks per second. If 10 machines are pipelined, each working on a different block, the latency remains 10 seconds, while the throughput rises to 1 block per second, allowing the system to keep up with the chain in a continuous, streaming manner.
    \paragraph{Proving Cost.} Often, zkVM proving is run on clusters of GPUs that are hosted in the cloud. These clusters are typically used on-demand, with GPUs that can be purchased instantaneously, or are reserved, via long-term agreements for compute. The \textit{proving cost} is the dollar cost of generating a ZK proof of a program given the underlying cluster on which proving occurs.     
    \subsubsection{Profiling Reth.} A common task for benchmarking the performance of zkVMs on real workloads is proving Ethereum block execution. Reth is an Ethereum client written in Rust \cite{reth2023} that can be used in conjunction with SP1 to prove EVM execution. This profiling enables a detailed understanding of the stages S0-S4 in which the zkVM spends the most time. In Table \ref{tab:reth-profiling}, we profile Ethereum proving with SP1 on various real blocks composed of different transaction types. For instance, block $21940007$ contains bn254 syscalls, which correspond to elliptic curve arithmetic, whereas block $21940004$ is a small block of roughly $16$ million RISC-V cycles. 
    
    We map the individual steps in proving Ethereum blocks to stages S0 to S4 from Figure \ref{fig:vAppSTF}. This mapping demonstrates that  Ethereum execution follows the standard vApp transaction lifecycle. We see that the zkVM spends a large fraction of its cycle count (approaching upwards of $63.31\%$ in some cases), in the block execution phase, corresponding to S2, the state transition function. For small blocks, only 28.96\% of cycles are spent on execution, as the cost of data access is not well amortized across a small number of transactions. We note that recover\_senders is optimized via an ECDSA recovery precompile. 
    
    In Table \ref{tab:interpreter-costs}, we show that executing programs inside the EVM can cause up to 832x cycle count overhead relative to simply executing native Rust in the zkVM. Combined with the transaction lifecycle profiling for Reth in Table \ref{tab:reth-profiling}, this allows us to realize opportunities for dramatic optimization. For instance, the block\_execution phase can be optimized with native Rust, and initialize\_witness\_db, which sets up an authenticated data structure for the account data and storage required for transaction execution, can be optimized with precompiles and specialized data structures \cite{qmdb}. Further, deserialize\_inputs can be optimized with a ZK friendly serializer.

\begin{table}[h]
\centering
\resizebox{\columnwidth}{!}{%
\begin{tabular}{lccc}
\toprule
\textbf{Block number}   & \textbf{21940007} & \textbf{21940014} & \textbf{21940004} \\
\textbf{Block type}     & bn254 syscalls    & bls12\_381 syscalls & small block      \\
\midrule
deserialize\_inputs (\textbf{S0})             & 12.66 \% & 16.67 \% & 22.47 \% \\
recover\_senders (\textbf{S0})                & 2.16 \%  & 3.08 \%  & 5.93 \%  \\
block\_validation (\textbf{S0} and \textbf{S4})  & 1.10 \%  & 1.27 \%  & 1.73 \%  \\
initialize\_witness\_db (\textbf{S1})         & 13.39 \% & 17.63 \% & 23.64 \% \\
block\_execution (\textbf{S2} and \textbf{S4})   & 63.31 \% & 51.29 \% & 28.96 \% \\
state\_root\_computation (\textbf{S3})        & 6.30 \%  & 8.72 \%  & 15.43 \% \\
accrue\_logs\_bloom (\textbf{S4})             & 0.52 \%  & 0.58 \%  & 0.68 \%  \\
other (\textbf{S0} and \textbf{S4})              & 0.56 \%  & 0.76 \%  & 1.15 \%  \\
\bottomrule
\end{tabular}%
}
\caption{Cycle counts by transaction lifecycle steps.}
\label{tab:reth-profiling}
\end{table}

    \subsubsection{Precompiles.}
    zkVMs benefit from optimization with \textit{precompiles}, which are custom circuits that are dedicated to proving particular operations. They accelerate the proving of commonly used primitives such as elliptic curve arithmetic, SHA256 hashing, and Keccak permutations. Instead of paying main CPU cycles and trace area for these operations, the zkVM instead dispatches the computations to their own tables and the CPU table of the zkVM looks up the appropriate values in these tables. Typically, writing precompiles for intensive operations can improve the performance of executing expensive operations in zkVMs by orders of magnitude. We illustrate the improvements in cycle counts from precompiles for various tasks in Table \ref{tab:precompile}. This table shows cycle count improvements for PlonK and Groth16 proof verification using a bn254 precompile for SP1; the precompile achieves upwards of $95\%$ improvement in cycle count. We also show improvements for KZG proof verification. This SP1 precompile achieves $95\%$ cycle count improvement for single proof verification and $78\%$ improvement for batch verification. The data contained in Table \ref{tab:precompile} was published in 
    \begin{center}
    \url{https://blog.succinct.xyz/succinctshipsprecompiles/}.
    \end{center}
\begin{table}[h]
\centering
\resizebox{\columnwidth}{!}{%
\begin{tabular}{lrrr}
\toprule
\multicolumn{4}{l}{\textbf{ZKP Systems}} \\
\midrule
\textbf{Operation} & \textbf{Before (cycles)} & \textbf{After (cycles)} & \textbf{\% Improvement} \\
PlonK Verification           & 187,227,852   & 8,078,761    & \textbf{95.69\%} \\
Groth16 Verification         & 173,953,261   & 9,390,640    & \textbf{94.60\%} \\
\addlinespace[0.5em]
\multicolumn{4}{l}{\textbf{KZG Proofs}} \\
\midrule
\textbf{Operation} & \textbf{Before (cycles)} & \textbf{After (cycles)} & \textbf{\% Improvement} \\
Verify KZG Proof                     & 212,709,402   & 9,391,832    & \textbf{95.58\%} \\
Verify Blob KZG Proof                & 265,322,934   & 27,960,797   & \textbf{89.46\%} \\
Verify Blob KZG Proof Batch (10×)    & 1,228,277,089 & 270,655,817  & \textbf{77.96\%} \\
\bottomrule
\end{tabular}%
}
\caption{Cycle count improvements after precompilation}
\label{tab:precompile}
\end{table}
\subsubsection{Repricing the EVM for zkVMs} Because the EVM was designed to emulate a CPU, its opcodes and precompiles may be dramatically mispriced relative to zkVMs. This was noted during zkEVM design considerations \cite{buterin2022zkevm} and in EIPs for Ethereum \cite{EIP7667}. zkVMs offer a unique opportunity to realign the EVM's costs with a cost model that closely approximates ZK proving. To demonstrate this mispricing, we conduct a precompile-level analysis for 600 Ethereum blocks. For each of these blocks, we compute the average RISC-V cycles spent on each EVM precompile by running the program in execution mode in SP1. This allows us to compute the average cycle count of the precompile in the zkVM. Table \ref{tab:mispricing} shows a comparison between the average cycle counts and the Ethereum gas of the various precompiles. These precompiles correspond to common cryptographic operations, such as sha256 (a hashing precompile), and modexp (which is used in RSA encryption). Our results show that certain precompiles, such as modexp and identity, are dramatically mispriced relative to their ZK proving costs. This data was collected using RSP, a repository that generates proofs of EVM execution using SP1, and is available at
\begin{center}
    \url{https://github.com/succinctlabs/rsp/}.
\end{center}

\begin{table}[H]
\centering
\resizebox{\columnwidth}{!}{%
\begin{tabular}{lcccc}
\toprule
\textbf{Precompile} & \textbf{Ethereum Gas} & \textbf{Avg Cycles} & \textbf{Std Cycles} & \textbf{Cycles / Gas} \\
\midrule
modexp                   & 200    & 173,033.7   & 348,068.8   & \cellcolor{red!80}865.2 \\
identity                 & 15     & 1,086.3     & 747.1       & \cellcolor{red!60}72.4 \\
bn-pair                  & 45,000 & 1,192,758.4 & 2,632,834.1 & \cellcolor{red!30}26.5 \\
sha256                   & 60     & 1,187.8     & 3,279.5     & \cellcolor{orange!30}19.8 \\
ecrecover                & 3,000  & 47,768.3    & 7,884.2     & \cellcolor{yellow!30}15.9 \\
bn-mul                   & 6,000  & 85,186.9    & 188,210.7   & \cellcolor{yellow!20}14.2 \\
bn-add                   & 150    & 1,553.1     & 3,376.2     & \cellcolor{green!20}10.4 \\
kzg-point-evaluation     & 50,000 & 392,191.9   & 1,890,609.4 & \cellcolor{green!40}7.8 \\
\bottomrule
\end{tabular}%
}
\caption{Cycle counts per Ethereum gas for EVM precompiles}
\label{tab:mispricing}
\end{table}
\subsubsection{Minimizing Instructions.} \label{chap:miniCode}zkVMs benefit from native Rust execution. Writing the state transition function of the application directly in Rust and compiling down to RISC-V reduces the overhead of passing the application through an interpreter like the EVM. This direct compilation to native RISC-V instructions eliminates the multiple translation layers required when processing EVM bytecode, resulting in significantly reduced computational overhead. Without the need for runtime interpretation, zkVMs achieve faster execution speeds and lower resource consumption, allowing for efficient generation of zero-knowledge proofs. Further, in the zkVM model, vApp developers can optimize their programs for a dramatically different cost profile than in the EVM. 

We illustrate this in Table \ref{tab:interpreter-costs}, where we show cycle counts and the overhead over native Rust for various programs in SP1. We compare Rust execution on three representative Fibonacci programs to Wasmi, which is a Web Assembly interpreter, and Revm, which is an interpreter for the EVM. Our results demonstrate that the EVM can cause up to 197x overhead even for benign programs like Fibonacci. This overhead occurs as a result of executing EVM-specific opcodes inside the zkVM. In the vApp framework, developers can write their applications directly in Rust and execute the program natively in the zkVM, avoiding interpreter overhead altogether. It also highlights the potential of rearchitecting the EVM using RISC-V to accelerate proof generation. The code to reproduce these experiments is available at 
\begin{center}
 \url{https://github.com/succinctlabs/overhead}.
\end{center}

\begin{table}[h]
\centering
\footnotesize
\renewcommand{\arraystretch}{1.1}
\begin{tabular}{p{4.2cm} r r}
\toprule
\textbf{Interpreter} & \textbf{Cycle Count} & \textbf{Overhead} \\
\midrule
\multicolumn{3}{l}{\textbf{Fibonacci-1k (u32)}} \\
Baseline (Native Rust) & 16,947        & 1x   \\
Wasmi                  & 347,000       & 20x  \\ 
\rowcolor{red!15} Revm (optimized Solidity) & 2,449,400     & 145x \\
\midrule
\multicolumn{3}{l}{\textbf{Fibonacci-10k (u64)}} \\
Baseline (Native Rust) & 124,947        & 1x   \\
Wasmi                  & 3,913,090      & 31x \\ 
\rowcolor{red!15} Revm (optimized Solidity) & 23,940,971     & 192x \\
\midrule
\multicolumn{3}{l}{\textbf{Fibonacci-100k (u128)}} \\
Baseline (Native Rust) & 1,204,947       & 1x   \\
Wasmi                  & 38,785,317      & 32x  \\ 
\rowcolor{red!15} Revm (optimized Solidity) & 237,603,026   & 197x \\
\bottomrule
\end{tabular}
\caption{Relative cycle count overhead of interpreters}
\label{tab:interpreter-costs}
\end{table}

\subsubsection{Parallelization: Sharding and Recursion}  
\begin{figure}[ht]
		\centering
        \setlength{\figWidth}{0.46\textwidth}
        \Description{Description of the image goes here.}
		\includegraphics[width=\figWidth]{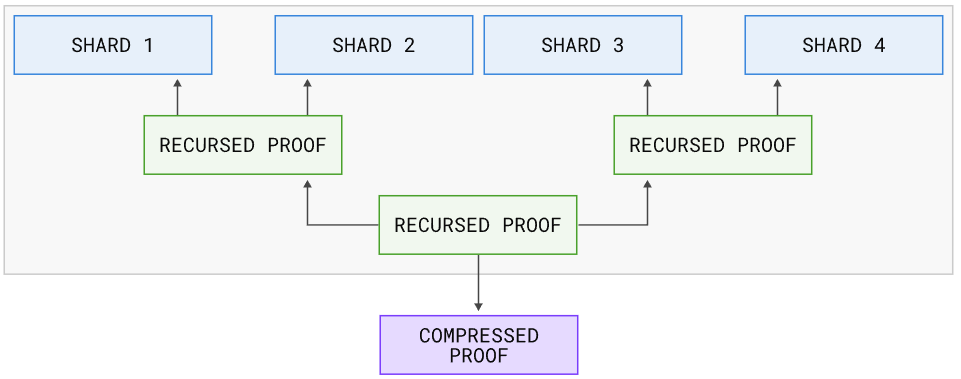}
		\caption{Sharding and recursion on a single machine can accelerate proving.}
        \label{fig:sharding-recursion}
\end{figure}

zkVMs benefit from parallelization at the level of a single machine. They can prove arbitrarily large programs by chunking the program's execution into multiple ``shards", which are contiguous batches of cycles, and generating proofs for each shard in parallel, with consistency checks across shards. The proofs corresponding to these shards can be combined together via a process called recursion. Recursion allows the zkVM to jointly verify proofs and compose them into a single, compressed proof. The large number of shard proofs can be recursed to create a small, efficiently verifiable proof, such as a Groth16 or PlonK proof. We note that not all proof systems currently support recursion. This is a form of parallelization that can be implemented on a CPU or accelerated hardware like GPUs, which makes zkVMs particularly suited to single-device optimization. We depict the process of sharding and recursion at the level of a proof on a single machine in Figure \ref{fig:sharding-recursion}.

\subsubsection{Parallelization: multi-GPU acceleration.} zkVMs are subject to even greater parallelization if proving is run on a cluster of many GPUs. Modern GPUs, with thousands of cores and high memory bandwidth, efficiently handle cryptographic operations such as Fast Fourier Transforms (FFTs) specific to ZK proving. Properly optimized zkVM implementations demonstrate near-linear scaling in throughput with additional GPUs. The ubiquitous availability of GPU clusters due to datacenter buildouts in related industries such as AI generates complementarities for zkVMs. Increasingly, production workloads for ZK rollups are being proven in clusters of thousands of GPUs and novel marketplaces for coordinating proving providers are emerging.

\subsubsection{Hardware Acceleration.} We conduct a study that demonstrates how recursion and GPU acceleration enable zkVMs to experience enhanced performance. In Table \ref{tab:ablation-setup}, we show the latency, throughput, and proof size of representative Ethereum block execution across a range of hardware configurations. We collect data for two representative blocks (20526624, a small block of 41.5 million cycles and 21141000, a large block of 278.2 million cycles); this data was collected on NVIDIA GeForce RTX 4090 GPU. We compare core proving (which involves only sharding, without recursion) on a CPU, with core proving on a GPU and compressed proving on a GPU (which includes recursion). We observe that GPU acceleration increases throughput over core CPU by up to 4x, and proof compression dramatically reduces proof sizes in large blocks by up to 300x, making these proofs suitable for onchain verification. This demonstrates the large performance improvements that are possible for zkVMs with hardware acceleration. Over time, optimized GPU implementations of novel proof systems and application-specific chips for ZK proving provide the possibility for ever-greater hardware acceleration.

\begin{table}[H]
\centering
\small
\renewcommand{\arraystretch}{1.1}
\begin{tabular}{
    l
    >{\raggedleft\arraybackslash}r
    >{\raggedleft\arraybackslash}r
    >{\raggedleft\arraybackslash}r
}
\toprule
\textbf{Setup} & \textbf{Latency} & \textbf{Throughput} & \textbf{Proof Size} \\
& \textbf{(s)} & \textbf{(KHz)} & \textbf{(MB)} \\
\midrule
\multicolumn{4}{l}{\textbf{Small Block (41.5M cycles)}} \\
CPU (core)       & 408.98    & 101.38   & 58.9 \\
GPU (core)       & 14.27    & 2906.29   & 58.9 \\
GPU (compressed)   & 30   & 1371.17   & 1.5 \\
\midrule
\multicolumn{4}{l}{\textbf{Big Block (278.2M cycles)}} \\
CPU (core)       & 2637.59  & 105.45   & 346\\
GPU (core)       & 88.7    & 3135.91   & 346\\
GPU (compressed)   & 145.91  & 1906.34  & 1.5 \\
\bottomrule
\end{tabular}
\caption{Experiment setup for hardware acceleration}
\label{tab:ablation-setup}
\end{table}

\subsection{Security Surface}
  
zkVMs are complex software systems that are composed of several modular components. They must withstand scrutiny across their architecture: from program compilation to proof generation. Their security rests on three key pillars: 

\begin{enumerate}[leftmargin=0pt,label={}]
    \item \textbf{(1) Soundness and completeness}: the underlying proof system (STARK or SNARK) must guarantee that a valid proof corresponds to a correct computation (soundness) and that every correct computation can produce a proof (completeness). STARKs, with no trusted setup, offer inherent transparency, while SNARKs require careful setup management.
    \item \textbf{(2) Implementation correctness}: Bugs in the compiler, runtime, or prover could undermine security. Open-sourcing the codebase (as is the case with several modern zkVMs) enables community audits. Formal verification, via mathematical proofs of system properties, and integration with secondary layers of security such as TEEs further reduces risks. 
    \item \textbf{(3) Modularity}: A modular design isolates components, making it easier to patch vulnerabilities or upgrade individual layers (e.g., swapping proof systems) without overhauling the entire stack.
\end{enumerate}
As zkVMs find their way in production systems via ZK rollups today and via vApps in the future, making their security surface increasingly transparent to application developers will become important. 
\section{Conclusion}
This paper introduced the vApp framework, realized through a developer friendly SDK in Rust. The framework supports the development of verifiable blockchain applications across diverse scenarios, including single-chain, inter-chain, and Web2-Web3 hybrid contexts, enabling rapid and secure integration of verifiable computing into Internet scale applications without extensive blockchain expertise.

From a developer's perspective, the SDK offers a universal codebase encompassing integrated macro constructs and libraries. This feature provides a high degree of abstraction in the generation of artifacts for verifiable blockchain applications, extending beyond typical rollup use cases.  This development environment enables application logic to be written once and deployed across various execution scenarios, supporting asynchronous and synchronous composability among different vApps, both within and across multiple blockchains.

Furthermore, benchmark results suggest that the integration of vApps with zkVMs can potentially achieve dramatic performance gains by utilizing specialized precompiles and hardware acceleration. Notably, cryptographic precompiles for operations such as PlonK and KZG proof verification reduced proving cycle counts by up to 95.7\%, while native Rust execution eliminates interpretation overhead, yielding up to 197x speedup compared to EVM-based approaches. GPU acceleration enhances proving throughput by up to 30x, and recursive proving compresses large proofs from 346MB to 1.5MB, enabling efficient on-chain verification and dramatically reducing latency and resource usage.

By dramatically lowering barriers to entry through composability, abstraction, and modularity, vApps simplify development and integration, paving the way for broader adoption of blockchain technology and setting the stage for a new generation of verifiable, internet-scale applications accessible to a wider range of developers.

	\begin{acks}
	We gratefully acknowledge the valuable feedback and assistance provided by individuals and teams regarding system design and benchmarking. We also extend our appreciation to Patrick O’Grady of Commonware, Ye Zhang of Scroll and Tarun Chitra of Robot Ventures for their insightful suggestions that improved the manuscript.
	\end{acks}
\balance
\bibliographystyle{unsrt}
\bibliography{bibliography}

\begin{thebibliography}{10}

\bibitem{scaling}
Ethereum.
\newblock Scaling.
\newblock \url{https://ethereum.org/en/developers/docs/scaling/}, 2025.

\bibitem{rollupsurvey}
{Louis Tremblay Thibault, Tom Sarry and Abdelhakim Senhaji Hafid}.
\newblock Blockchain scaling using rollups: A comprehensive survey.
\newblock IEEE Access, vol. 10, 93039-93054, 2022.

\bibitem{snark}
Bryan Parno, Craig Gentry, Jon Howell, and Mariana Raykova.
\newblock Pinocchio: Nearly practical verifiable computation.
\newblock \url{https://eprint.iacr.org/2013/279}, 2013.

\bibitem{stark}
Eli Ben-Sasson, Iddo Bentov, Yinon Horesh, and Michael Riabzev.
\newblock Scalable, transparent, and post-quantum secure computational integrity.
\newblock \url{https://eprint.iacr.org/2018/046}, 2018.

\bibitem{cryptoTEE}
Victor Costan and Srinivas Devadas.
\newblock Intel {SGX} explained.
\newblock \url{https://eprint.iacr.org/2016/086}, 2016.

\bibitem{optrollup}
OffchainLabs.
\newblock {Offchainlabs}.
\newblock \url{https://github.com/OffchainLabs/nitro/blob/master/docs/Nitro-whitepaper.pdf}, 2022.

\bibitem{codeFrag}
Stefanos Chaliasos, Arthur Gervais, and Benjamin Livshits.
\newblock A study of inline assembly in solidity smart contracts.
\newblock Proceedings of the ACM on Programming Languages 6.OOPSLA2 (2022): 1123-1149, 2022.

\bibitem{layerzero}
Ryan Zarick, Bryan Pellegrino, Isaac Zhang, Thomas Kim, and Caleb Banister.
\newblock Layerzero.
\newblock \url{https://arxiv.org/abs/2312.09118}, 2024.

\bibitem{atomSwap}
Léonard Lys, Arthur Micoulet, and Maria Potop-Butucaru.
\newblock R-{SWAP}: Relay based atomic cross-chain swap protocol.
\newblock \url{https://eprint.iacr.org/2021/621}, 2021.

\bibitem{graphQLBook}
GraphQL.
\newblock {GraphQL}.
\newblock \url{https://spec.graphql.org/October2021/}, 2021.

\bibitem{qmdb}
Isaac Zhang, Ryan Zarick, Daniel Wong, Thomas Kim, Bryan Pellegrino, Mignon Li, and Kelvin Wong.
\newblock Qmdb: Quick merkle database.
\newblock \url{https://arxiv.org/abs/2501.05262}, 2025.

\bibitem{RecurZKP}
Sean Bowe, Jack Grigg, and Daira Hopwood.
\newblock Recursive proof composition without a trusted setup.
\newblock \url{https://eprint.iacr.org/2019/1021}, 2019.

\bibitem{PCD}
Nir Bitansky, Ran Canetti, Alessandro Chiesa, and Eran Tromer.
\newblock Recursive composition and bootstrapping for {SNARKs} and proof-carrying data.
\newblock \url{https://eprint.iacr.org/2012/095}, 2013.

\bibitem{blsSig}
Dan Boneh, Ben Lynn, and Hovav Shacham.
\newblock Short signatures from the weil pairing.
\newblock Journal of Cryptology, vol. 17, 297-319, 2004.

\bibitem{buterin2022zkevm}
Vitalik Buterin.
\newblock The different types of zk-evms.
\newblock \url{https://vitalik.eth.limo/general/2022/08/04/zkevm.html}, August 2022.

\bibitem{succinct_sp1}
Succinct.
\newblock {SP1}.
\newblock \url{https://github.com/succinctlabs/sp1}, 2025.

\bibitem{risc0}
RiscZero.
\newblock risc0.
\newblock \url{https://github.com/risc0/risc0}, 2025.

\bibitem{openvm}
OpenVM.
\newblock Openvm: A performant and modular zkvm framework built for customization and extensibility.
\newblock \url{https://github.com/openvm-org/openvm}, 2025.

\bibitem{a16z-jolt}
a16z.
\newblock Jolt: The simplest and most extensible zkvm.
\newblock \url{https://github.com/a16z/jolt}.

\bibitem{mantle2024zkrollup}
Mantle.
\newblock Mantle network advances technical roadmap as the first zk validity rollup with succinct’s sp1.
\newblock \url{https://www.mantle.xyz/blog/announcements/mantle-network-advances-technical-roadmap-as-the-first-zk-validity-rollup-with-succincts-sp1}, 2024.

\bibitem{sp1-blobstream}
Succinct Labs.
\newblock Sp1 blobstream.
\newblock \url{https://github.com/succinctlabs/sp1-blobstream}, 2025.
\newblock Implementation of zero-knowledge proof circuits for Blobstream, Celestia's data attestation bridge, in SP1.

\bibitem{sp1-vector}
Succinct Labs.
\newblock Sp1 vector.
\newblock \url{https://github.com/succinctlabs/sp1-vector}, 2025.
\newblock Implementation of zero-knowledge proof circuits for Vector, Avail's data attestation bridge, in SP1.

\bibitem{taiko2024basedrollup}
Taiko Team.
\newblock Based rollup: A new type of ethereum rollup.
\newblock \url{https://taiko.mirror.xyz/_3RbETXwvtYmK0T8j7VOllPw2DXT4gt_e54qqs8V-Lc}, 2024.

\bibitem{gabizon2019plonk}
Ariel Gabizon, Zachary Williamson, and Oana Ciobotaru.
\newblock Plonk: Permutations over lagrange-bases for oecumenical noninteractive arguments of knowledge.
\newblock \url{https://eprint.iacr.org/2019/953}, 2019.
\newblock Report 2019/953.

\bibitem{groth2016size}
Jens Groth.
\newblock On the size of pairing-based non-interactive arguments.
\newblock \url{https://eprint.iacr.org/2016/260}, 2016.

\bibitem{reth2023}
{Paradigm}.
\newblock Reth: A next-generation ethereum client.
\newblock \url{https://github.com/paradigmxyz/reth}, 2023.

\bibitem{EIP7667}
Vitalik Buterin.
\newblock {EIP-7667: Raise gas costs of hash functions}.
\newblock \url{https://eips.ethereum.org/EIPS/eip-7667}, 2024.

\end{thebibliography}
\end{document}